\documentclass[superscriptaddress,aps,prd,reprint,nofootinbib]{revtex4-2}

\usepackage{lipsum}
\usepackage{blindtext}
\usepackage{graphicx}

\usepackage{amssymb} \usepackage{amsmath} 
\usepackage{epsfig,latexsym}
\usepackage{dsfont}

\usepackage{dcolumn}
\usepackage{bm}

\usepackage{ulem}
\usepackage{slashed}
\usepackage{amsfonts}
\usepackage{mathrsfs}
\usepackage{color}
\usepackage[makeroom]{cancel}
\usepackage{graphicx}
\usepackage{amssymb}
\usepackage{amsmath}
\usepackage{hyperref}
\usepackage{cleveref}
\usepackage{hyperref}
\renewcommand{\d}{\textrm{d}}
\newcommand{\e}{\textrm{e}}

\newcommand{\RNum}[1]{\uppercase\expandafter{\romannumeral #1\relax}}

\def\bi{\begin{itemize}}
\def\ei{\end{itemize}}
\def\be{\begin{equation}}
\def\ee{\end{equation}}
\newcommand{\bea}{\begin{eqnarray}}
\newcommand{\eea}{\end{eqnarray}}

\newcommand{\red}{\textcolor{red}}

\begin{document}

\preprint{APS/123-QED}

\title{Can We Probe Spacetime Non-commutativity Through Tidal Deformability of Compact Objects?}

\author{Junting Peng}
\affiliation{Center for Field Theory and Particle Physics \& Department of Physics, Fudan University, 200433 Shanghai, China}

\author{Yanbo Zhao}
\affiliation{Center for Field Theory and Particle Physics \& Department of Physics, Fudan University, 200433 Shanghai, China}

\author{Antonino Marcian\`o}
\email{marciano@fudan.edu.cn}
\affiliation{Center for Field Theory and Particle Physics \& Department of Physics, Fudan University, 200433 Shanghai, China}
\affiliation{Laboratori Nazionali di Frascati INFN, Frascati (Rome), Italy, EU}
\affiliation{INFN sezione Roma Tor Vergata, I-00133 Rome, Italy, EU}

\date{\today}

\begin{abstract}
\noindent
  We investigate the impact of spacetime non-commutativity on the tidal deformability of compact objects and explore the feasibility of detecting non-commutative (NC) effects through gravitational wave (GW) observations. We considered NC modifications to spacetime geometry based on de Sitter gauge theory of gravity and calculate their impact on tidal deformability. While several types of compact objects have been proposed as candidates for probing spacetime non-commutativity, particularly at the horizon scales, our study showed analytically that,  for compact objects with non-singular metric at their surface (such as neutron stars and boson stars), the NC correction to their tidal deformability converge to a finite value at the black-hole-compactness limit, eliminating infinite enhancement at the horizon scales. We then compute the NC corrections for neutron stars and boson stars, considering several different models, and analyze their imprints on the GW signals. By comparing the results, we assess the scale of NC effects across different compactness regimes and discuss the conditions under which these NC effects can be amplified. While our findings suggest that the leading-order NC correction dominates the tidal deformability of a compact object near the black-hole-compactness limit, we demonstrate that neutron stars and boson stars are not viable candidates to constrain spacetime non-commutativity, while relying on the tidal deformability through GW observations.
\end{abstract}

\maketitle

\section{Introduction}\label{1}
\noindent
Spacetime non-commutativity was first introduced as a framework to quantize spacetime while preserving Lorentz invariance or modified Lorentz invariance \cite{Snyder:1946qz}. In later decades, it was explored within several approaches to quantum gravity \cite{Connes_1985, chamseddine1993, Connes_1998, Witten:1985cc, Seiberg_1999, Kanazawa}. In particular, N.~Seiberg and E.~Witten developed the Seiberg-Witten map, a method for computing non-commutative (NC) corrections to Yang-Mills gauge field theories \cite{Seiberg_1999}. Among many other implications, this approach provided a novel path to incorporate NC corrections into the metric field, in light of multiple attempts to reinterpret gravity as a gauge theory \cite{M.Agop_2008, V.Enache_2008, Chiritoiu, Zet}. Notably, A.H.~Chamseddine derived analytical NC corrections to tetrads using the Seiberg-Witten map within the de Sitter gauge theory of gravity \cite{Chamseddine_2001}. Efforts have also been made to calculate NC effects on observable quantities based on corrections to the Schwarzschild metric  \cite{Chaichian_2008, Kanazawa, Zhao_2024}.

We discuss here the possibility of constraining spacetime non-commutativity through tidal deformability, the latter depending on the spacetime metric both inside and outside celestial bodies \cite{Hinderer_2010, Hinderer_2008}. NC effects are also supposed to imprint the phase of gravitational wave (GW) signals emitted by binary systems \cite{Flanagan_2008}.

We first derive NC corrections to the tidal deformability of general spherical compact objects with non-singular metric on their surface. We hence analyze the NC corrections to GW signals emitted by binary inspiraling events that involve these compact objects. As we will emphasize in the following sections, we found for the NC corrections a finite behavior in the black-hole-compactness limit. This prevents any ultra-Planckian enhancement in space localization due to NC effects, when the radius of the object approaches the horizon scale~\cite{Maselli_Horizon_Scale}.

To obtain numerical results, we begin by focusing on neutron stars, for which tidal deformability calculations require a detailed consideration of the equation of state (EoS). We focus on the two cases of the SLy4 EoS \cite{Douchin_2001} and the FPS EoS \cite{Pandharipande1989}. The EoS data are obtained from the work of P.~Haensel and A.~Y.~Potekhin \cite{Haensel_2004} and interpolated for numerical calculations. We hence evaluate NC effects on tidal deformability for the chosen neutron star EoSs, along with their impact on the GW phase, for different values of the binary neutron star masses. 

We then extend the analysis to boson stars, hypothetical exotic compact objects\footnote{For instance, it has been suggested that scalar axions could form stable boson stars \cite{Edward_axion_BS}.} with uncertain EoS. There exist several studies that concern the boson stars structure, which are based on scalar field theory \cite{Cardoso_TL_BS}\cite{Mecedo_BS}. We addressed here the boson star structure via a specific algorithm based on semiclassical methodologies, and further analyzed the NC effects on boson stars and boson star binaries, including the limit of these effects when the star surfaces are close to the horizon scale.

By combining analytical and numerical results, we discuss the scale of NC corrections to the tidal Love number of compact objects and their impact on GW phase shifts across different scenarios and discuss the conditions under which these NC effects can be amplified. Finally, we assess the feasibility of detecting NC effects through GW observations of compact binary mergers.

The structure of the paper is as follows. Section~\ref{2} reviews NC corrections to spacetime metrics. Section~\ref{3} analytically derives NC modifications on tidal deformability and GW signals, as well as their results in the black hole-compactness limit. Section~\ref{4} presents the numerical results for neutron stars. Section~\ref{5} presents numerical results for boson stars and discusses the feasibility of observing these NC effects across different scenarios. Section~\ref{6} we spell out the conclusions. In the paper, all units are Planckian unless otherwise specified.

\section{Corrections to the metric from
Spacetime non-commutativity}\label{2}
\subsection{Gauge Theory of Gravity}
\noindent
To calculate the corrections to the background metric induced by spacetime non-commutativity, we first review the de Sitter gauge theory of gravity. The metric of a four-dimensional curved spacetime can be introduced as a component of an \(\text{SO}(4,1)\) gauge field ~\cite{Zet}. This gauge field is a connection one-form with ten independent internal components, which transform under the adjoint representation of \(\text{SO}(4,1)\).

We denote the components of the \(\text{SO}(4,1)\) gauge field as \(\omega^{AB}_\mu\), which satisfy the anti-symmetry condition:
\begin{equation}
    \omega^{AB}_\mu(x)=-\omega^{BA}_\mu(x),
\end{equation}
where \(A, B = 0, 1, 2, 3, 4\) are indices in the fundamental representation of the group. We denote by lower-case Latin letter a subset of these indices, namely \(a,b = 0, 1, 2, 3\), such that the components \(\omega^{ab}_\mu(x)\) correspond to the six generators that transform under the adjoint representation of the Lorentz group. 

The remaining components of the \(\text{SO}(4,1)\) connection can be identified as \(\omega^{a5}_\mu(x) = k e^a_\mu(x)\), where \(k\) is a constant, and \(e^a_\mu(x)\) represents the tetrad field:
\begin{equation}
    g_{\mu\nu}=\eta_{ab}e^a_\mu e^b_{\nu} \,,\label{Tetrad}
\end{equation}
with $\eta_{ab}={\rm diag}\{-1, 1 , 1 , 1\}$ Minkowski spacetime metric. \\

The field strength of the SO(4,1) connection reads
\begin{eqnarray}
    F^{AB}_{\mu\nu}&=&\partial_\mu \omega^{AB}_\nu-\partial_\nu \omega^{AB}_\mu \nonumber \\
    &\phantom{a}& + f^{AB}_{\ \  \ \  CD \, EF}\,\omega^{CD}_\mu \,\omega^{EF}_\nu\,,
\end{eqnarray}
where $f^{AB}_{\ \  \ \ CD \, EF}$ are the structure constants of the SO(4,1) group, expressed in terms of the flat internal metric $\eta_{AB}$, i.e.
\begin{eqnarray} \label{str}
f^{AB\ LM\ PQ} &\equiv& \frac{1}{4}(\eta^{AL} \eta^{BQ} \eta^{MP} - \eta^{BL} \eta^{AQ} \eta^{MP} \nonumber\\ 
&-& \eta^{AM} \eta^{BQ} \eta^{LP} + \eta^{BM}  \eta^{AQ} \eta^{LP} \nonumber\\ 
&-& P\longleftrightarrow Q)\,,    
\end{eqnarray} 
with $\eta_{AB}={\rm diag}\{-1, 1 , 1 , 1, 1\}$. Exploiting this result, the Riemann curvature tensor and the torsion tensor can be introduced to be, respectively,
\begin{align}
    F^{ab}_{\mu\nu}=&\partial_\mu \omega^{ab}_\nu-\partial_\nu \omega^{ab}_\mu+(\omega^{ac}_\mu \omega^{db}_\nu-\omega^{ac}_\nu \omega^{db}_\mu) \eta_{cd}
    \nonumber\\
    &+k^2(e^a_\mu  e^b_\nu-e^a_\nu  e^b_\mu)\nonumber\\
    :=&R^{ab}_{\mu\nu}\,,
    \label{FL}
\end{align}
\begin{align}
	F^{a5}_{\mu\nu}=&k[\partial_\mu e^{a}_\nu-\partial_\nu e^{a}_\mu+(\omega^{ab}_\mu  e^{c}_\nu-\omega^{ab}_\nu  e^{c}_\mu)\eta_{bc}]\nonumber\\:=& k T^a_{\mu\nu} \,.
     \label{deffs}
\end{align}
In Eq.~\eqref{RiemannT} we have introduced the definition of the Riemann tensor with mixed indices. This is related to the standard expression of the Riemann tensor by contracting the internal indices with the components of the tetrad and its inverse  
\begin{equation}
R^\sigma_{\mu\nu\rho}=R^{ab}_{\mu\nu} e^\lambda_a e^\sigma_b g_{\lambda\rho}\,,
\label{RiemannT}
\end{equation}
where $e^\lambda_a$ denotes the inverse of $e^a_\lambda$, defined by  $e^\mu_a e^a_\nu=\delta^\mu_\nu$. \\

In what follows we will consider torsion to be absent, namely  
\begin{equation} \label{cart}
T^a_{\mu\nu}=0\,,
\end{equation}
which is nothing but the Cartan structure equation --- namely, the gauge Gauss constraint for SO(3,1) --- expressing the metric compatibility of the gravitational spin-connection $\omega_\mu^{ab}$. The solution to Eq.~\eqref{cart} thus enables to derive the torsion-free contribution to the spin connection in terms of vierbein field $e^a_\mu$, i.e.
\begin{equation}
    \begin{aligned}
        \omega^{ab}_\mu[e]=&\frac12e^\nu_c(\partial_\mu e^b_\nu-\partial_\nu e^b_\mu)\eta^{ac}-\frac12e^\nu_c(\partial_\mu e^a_\nu-\partial_\nu e^a_\mu)\eta^{bc}\\
        &-\frac12e^\rho_d e^\sigma_m(\partial_\rho e_{\sigma c}-\partial_\sigma e_{\rho c})e^c_\mu \eta^{ad} \eta^{bm}\,.
        \label{SC}
    \end{aligned}
\end{equation}

We may now consider the $k\rightarrow0$ limit, for which the definition of $R^{\sigma}_{\mu\nu\rho}$ is totally equivalent to the definition of the Riemann curvature in differential geometry. The Einstein field equations are hence satisfied, namely
\begin{equation}
    R_{\mu\nu}-\frac{1}{2}g_{\mu\nu}R=0\,,
\end{equation}
where $R_{\mu\nu}$ is the Ricci tensor and $R$ is the scalar curvature.

\subsection{Non-Commutative Corrections to the Metric}
\noindent
The starting point for our discussion on spacetime non-commutativity is the commutator among spacetime coordinates, which for the Moyal plane reads
\begin{equation}
    [x^\mu, x^\nu] = i \Theta^{\mu\nu}\,.
\end{equation}

Adopting the Seiberg-Witten map \cite{Seiberg_1999}, non-commutative quantities can be mapped into commutative one. This happens thanks to a non-commutative algebra that is introduced through the Moyal product, defined by
\begin{equation}
    f(x)*g(x):=e^{\frac{i}{2}\Theta^{\mu\nu}\frac{\partial}{\partial\xi^\mu}\frac{\partial}{\partial\zeta^\mu}}f(x+\xi)g(y+\zeta)|_{\xi=\zeta=0}\,,
    \label{MY}
\end{equation}
where $f(x)$ and $g(x)$ are functions of the spacetime manifold. 

Applying the Seiberg-Witten map to relate the $\text{SO}(4,1)$ gauge fields to their non-commutative corrections, $e^a_\mu$ can be evaluated up to the second order in $\Theta^{\mu\nu}$~\cite{Chamseddine_2001,Chaichian_2008}, i.e.
\begin{equation}
\scalebox{0.8}{$
\begin{aligned}
    \hat{e}_\mu^a(x, \Theta)=e_\mu^a(x)-i \Theta^{\nu \rho} e_{\mu \nu \rho}^a(x)+\Theta^{\nu \rho} \Theta^{\lambda \tau} e_{\mu \nu \rho \lambda \tau}^a(x)+O\left(\Theta^3\right)\,,
\end{aligned}
$}
\label{e_expansion}
\end{equation}
where
\begin{equation}
\scalebox{0.8}{$
\begin{aligned}
& e_{\mu \nu \rho}^a=\frac{1}{4}\left[\omega_\nu^{a c} \partial_\rho e_\mu^d+\left(\partial_\rho \omega_\mu^{a c}+R_{\rho \mu}^{a c}\right) e_\nu^d\right] \eta_{c d}, 
\end{aligned}
$}
\label{e first}
\end{equation}
\begin{equation}
\scalebox{0.8}{$
\begin{aligned}
    & e_{\mu \nu \rho \lambda \tau}^a=\frac{1}{32}[2\{R_{\tau \nu}, R_{\mu \rho}\}^{a b} e_\lambda^c-\omega_\lambda^{a b}(D_\rho R_{\tau \mu}^{c d}+\partial_\rho R_{\tau \mu}^{c d}) e_\nu^m \eta_{d m} \\
    &-\{\omega_\nu,(D_\rho R_{\tau \mu}+\partial_\rho R_{\tau \mu})\}^{a b} e_\lambda^c-\partial_\tau\{\omega_\nu,(\partial_\rho \omega_\mu+R_{\rho \mu})\}^{a b} e_\lambda^c \\
    &-\omega_\lambda^{a b} \partial_\tau(\omega_\nu^{c d} \partial_\rho e_\mu^m+(\partial_\rho \omega_\mu^{c d}+R_{\rho \mu}^{c d}) e_\nu^m) \eta_{d m}+2 \partial_\nu \omega_\lambda^{a b} \partial_\rho \partial_\tau e_\mu^c \\
    &-2 \partial_\rho(\partial_\tau \omega_\mu^{a b}+R_{\tau \mu}^{a b}) \partial_\nu e_\lambda^c-\{\omega_\nu,(\partial_\rho \omega_\lambda+R_{\rho \lambda})\}^{a b} \partial_\tau e_\mu^c \\
    &.-(\partial_\tau \omega_\mu^{a b}+R_{\tau \mu}^{a b})(\omega_\nu^{c d} \partial_\rho e_\lambda^m+(\partial_\rho \omega_\lambda^{c d}+R_{\rho \lambda}^{c d}) e_\nu^m \eta_{d m})] \eta_{b c}\,,
\end{aligned}
$}
\label{e second}
\end{equation}
the curly brackets referring to anti-commutators in the Latin indices, i.e. $\{\alpha, \beta\}^{ab}=\eta_{cd}(\alpha^{ac}\beta^{db}+\beta^{ac}\alpha^{db})$.\\

While the uncorrected metric $g_{\mu\nu}$ is related to tetrad base field $e^a_\mu$ by Eq.~\eqref{Tetrad}, the expression of the metric that encodes non-commutative corrections is similarly related to $\hat e^a_\mu$ by
\begin{equation}
	\hat{g}_{\mu \nu}(x, \Theta)=\frac{1}{2} \eta_{a b}\left(\hat{e}_\mu^a * \hat{e}_\nu^{b+}+\hat{e}_\nu^b * \hat{e}_\mu^{a+}\right)\,.
    \label{metric}
\end{equation}
This relations ensures that the metric is real and symmetric. Exploiting Eqs.~\eqref{e_expansion}, \eqref{e first}, \eqref{e second} and \eqref{metric}, it is possible to calculate the non-commutative corrections to the metric up to the second order in $\Theta^{\mu\nu}$.

\section{Testing Spacetime Non-commutativity with Tidal deformability}\label{3}

\subsection{Tidal Deformability and Gravitational Waves}
\noindent
Tidal deformability describes up to what extent an extended body can be deformed, once it is displaced in an external tidal field. In the inspiraling phase of a binary system, the tidal deformabilities of both the two stars will be imprinted in the gravitational wave signal they generate as a phase term \cite{Flanagan_2008} \cite{Hinderer_2010}. Meanwhile, the tidal deformability is closely connected to the metric inside and outside the body. Thus NC corrections to the metric field may in principle affect the tidal deformability, and eventually become manifest in GW signals. 

The background spacetime of a spherical object can be described in terms of a spherically symmetric metric, i.e.
\begin{equation}
    ds^2=-\e^{2\Gamma(r)}dt^2+\e^{2\Lambda(r)}dr^2+r^2d\theta^2+r^2\sin^2\theta d\varphi^2\,.
\end{equation}

The tidal deformability is calculated by considering a perturbation of the background metric of the form
\begin{equation}
    \begin{aligned}
        ds^2=&-\e^{2\Gamma(r)}[1+H(r)Y_{20}(\theta, \varphi)]dt^2\\&+\e^{2\Lambda(r)}[1-H(r)Y_{20}(\theta, \varphi)]dr^2\\&+r^2[1-K(r)Y_{20}(\theta, \varphi)](d\theta^2+\sin^2\theta d\varphi^2)\,.
    \end{aligned} \label{DM}
\end{equation} 

This induces a perturbed expression at the linear order of the Einstein field equations $R^\mu_\nu-\frac{1}{2}\delta^\mu_\nu R = 0$.
Outside the spherical object, this expression provides
{\small
\begin{align}
    H^{\prime\prime}(r)\!+\!\frac{2(r-M)}{r(r-2M)}H^\prime(r)\!-\!\frac{6r^2\!-\!12rM\!+\!4M^2}{r^2(r-2M)^2}H(r)\!=\!0, \label{HDEex}
\end{align}}
{\small
\begin{align}
    K^\prime(r)=H^\prime(r)+\frac{2M}{r(r-2M)} H(r)\,, \label{KDEex}
\end{align}
}where $M$ is the total mass of the star. Changing radial variable according to $X={r}/{M}-1$, Eq.~\eqref{HDEex} turns into a Legendre equation (with l=2, m=2), the general solution of which reads
{\small
\begin{align}
    H(r)=&c_1P^2_2(\frac{r}{M}-1)+c_2Q^2_2(\frac{r}{M}-1) \label{Hout}\\
    =&3(\frac{r}{M})^2c_1-6(\frac{r}{M})c_1+\frac{8}{5}(\frac{M}{r})^3 c_2+O((\frac{M}{r})^4) \,,\label{HExpan}
\end{align}
where $P^2_2$ and $Q^2_2$ are Legendre function of the first and of the second kind for $l=2, m=2$, while $c_1$ and $c_2$ are undetermined constants.\\

We may then write the effective Newtonian potential of the metric outside the star, and perform the multipole expansion
\begin{align}
        V_{\rm eff}=&-\frac{1+g_{00}}{2} \label{Veff}\\
		=&-\frac{M}{r}+\frac{1}{2}H(r)Y_{20}(\theta,\phi)(1-\frac{2M}{r}) \label{effH}\\
		=&-\frac{M}{r}-\frac{3}{2r^3}QY_{20}(\theta,\phi)+O(\frac{1}{r^4})\nonumber\\&+f_i n^i r+\frac{\epsilon Y_{20}(\theta,\phi)}{2}r^2+O(r^3) \,,\label{QEp}
\end{align}
where $Q$ and $\epsilon$ represent the tidal deformation and the tidal field strength, respectively, and $n^i=\frac{x^i}{r}$ is the direction vector. The tidal deformability is defined as $\lambda:=-\frac{Q}{\epsilon}$, i.e. the ratio between the tidal deformation and the strength of the external tidal field. It can be calculated by substituting Eq.~\eqref{HExpan} into Eq.~\eqref{effH}, and comparing the result with Eq.~\eqref{QEp}, namely
\begin{equation}
    \lambda:=-\frac{Q}{\epsilon}=\frac{8 M^5}{45}\frac{c_2}{c_1}\,,
\end{equation}
in which $c_1$ and $c_2$ can be calculated by solving numerically the perturbed Einstein field equations. These are differential equations for $H(r)$ and $K(r)$ inside the spherical body. Inserting the continuity of $H(r)$ and $H'(r)$ at $r=R$, one finds the expressions 
\begin{equation}
    \left\{
    \begin{aligned}
        &c_1P^2_2(\frac{R}{M}-1)+c_2Q^2_2(\frac{R}{M}-1)=H(R)\,,\\
        &\frac{c_1}{M}{P^2_2}^\prime(\frac{R}{M}-1)+\frac{c_2}{M}{Q^2_2}^\prime(\frac{R}{M}-1)=H'(R)\,.
    \end{aligned}
    \label{cSolve}
    \right.
\end{equation}

For the numerical calculations of $H(r)$ and $K(r)$, it is very useful to consider the substitutions 
\begin{align}
    h(r) :=& \frac{r H'(r)}{H(r)},\label{hdef}\\
    k(r) :=& \frac{K(r)}{H(r)},\label{kdef}
\end{align}
where $h(r)$ and $k(r)$ always encode well-behaved solutions for compact objects like neutron stars and boson stars, as will be discussed in the following sections.\\

To introduce the concept of tidal deformability, it is convenient to use the dimensionless tidal Love number 
\begin{equation}
k_2=\frac{3}{2}\lambda R^{-5}=\frac{4 c_2 M^5}{15 c_1 R^5}\,.
\end{equation}
By solving Eq.~\eqref{cSolve}, and consider the substitution of Eq.~\eqref{hdef}, its expression reads
\begin{equation}
\scalebox{0.9}{$
\begin{aligned}
    k_2 =& 8C^5(1-2C)^2[2(1-C)-h(R)(1-2C)]\\
    &\{2C(4C^4+6C^3-22C^2+15C-3)h(R)\\
    &+4C(2C^4-2C^3+13C^2-12C+3)\\
    &-3(1-2C)^2[h(R)(1-2C)-2(1-C)]\ln(1-2C)\}^{-1}\,,
\end{aligned}
$}
\end{equation}
where $C = {M}/{R}$ represents the compactness of the body. We can infer from this expression that in the black-hole-compactness limit, namely, when $C\rightarrow\frac{1}{2}$, $k_2$ will approach 0. \\

For the GWs emitted by an inspiraling binary system, we consider the Fourier transformation $\tilde{h}(f)=\mathcal{A}(f)\exp(i\Psi(f))$, with $f$ GW frequency. The phase $\Psi(f)$ contains a term sensitive to the tidal deformability of the two component stars in the binary system \cite{Flanagan_2008} \cite{Hinderer_2010}. This term, denoted as $\Psi_{TD}$, is expressed (in SI units) by the formula
\begin{equation}
\scalebox{0.8}{$
    \Psi_{TD} \!=\! -\,\frac{9}{16}\frac{[\pi (m_1+m_2) f]^\frac{5}{3} c^5}{\mu (m_1+m_2)^4 G^\frac{10}{3}}\Bigg[\left(12 \frac{m_2}{m_1} + 1\right) \lambda_1
    + \left(12 \frac{m_1}{m_2} + 1\right) \lambda_2\Bigg] \,,
$}
\label{PsiTD}
\end{equation}
where the subscripts $1$ and $2$ refer to the two component stars, $m$ is the mass of the star, $\lambda$ is the tidal deformability of the star, and finally the reduced mass is denoted by $\mu = \frac{m_1 m_2}{m_1 + m_2}$. Thus, Eq.~\eqref{PsiTD} is instrumental to determine the tidal deformability of the component stars from GWs observation~\cite{Chatziioannou_2020}.

\subsection{Non-Commutative Corrections to Tidal Deformability and Black-Hole-Compactness Limit}
\label{sec:k2cr_NS}
\noindent
 Following the methodology introduced in Section~\ref{2}, we calculate the spacetime NC corrections to the tidal deformability. A deformed spacetime metric near the spherical object, as appears in Eq.~\eqref{DM}, can be expressed by Eq.~\eqref{Tetrad} using a tetrad field $e^a_\mu$ of the form:
 \begin{equation}
     \begin{aligned}
	&e^0_\mu=\{e^\Gamma(r)\sqrt{1+H(r)Y_{20}(\theta,\phi)},0,0,0\}\,,\\
	&e^1_\mu=\{0,e^\Lambda(r)\sqrt{1-H(r)Y_{20}(\theta,\phi)},0,0\}\,,\\
	&e^2_\mu=\{0,0,r\sqrt{1-K(r)Y_{20}(\theta,\phi)},0\}\,,\\
	&e^3_\mu=\{0,0,0,r\sin\theta\sqrt{1-K(r)Y_{20}(\theta,\phi)}\}\,.
	\end{aligned}
 \end{equation}
 
 Besides, we consider a simplified model, whose coordinates' system only involves one non-vanishing parameter of non-commutativity: $\Theta^{r\theta}=-\Theta^{\theta r}=\Theta$ --- see e.g. Refs.~\cite{Chaichian_2008} \cite{Kanazawa}  \cite{Zhao_2024} --- the other components of $\Theta^{\mu\nu}$ being all vanishing \footnote{Row and column indices are both arranged following the order $t, r, \theta, \varphi$.}, namely 
\begin{equation}
    \Theta^{\mu\nu}=\left(\begin{array}{cccc}
	0 & 0 & 0 & 0 \\
	0 & 0 & \Theta & 0 \\
	0 & -\Theta & 0 & 0 \\
	0 & 0 & 0 & 0
	\end{array}\right)\,.
    \label{Theta model}
\end{equation}
$\Theta$ has dimension of length $L$, being proportional to the commutator $[r,\theta]$, rather than $L^2$, as it happens for the components of $\Theta^{\mu\nu}$ in the Cartesian coordinates' system. \\

We can now use Eqs.~\eqref{e_expansion}-\eqref{metric}, in order to derive the NC correction to the effective potential $V_{\rm eff}$, as defined in Eq.~\eqref{Veff}, up to the linear order in $H(r)$ and $K(r)$, namely 
\begin{equation}
    \scalebox{0.9}{$
	\begin{aligned}
	V_{\rm eff}&=-\frac{M}{r}+\frac{1}{2}H(r)Y_{20}(\theta,\phi)(1-\frac{2M}{r})+\frac{M(4r-11M)}{8r^4}\Theta^2 \\&+\frac{\Theta^2}{64r^4(r-2M)}\bigg\{\Big[4 M (89 M^2 - 80 M r + 16 r^2) H(r) \\&- 
	8 M (22 M^2 - 19 M r + 4 r^2) K(r) \\&-(2 M - r) r \big[2 M (57 M - 13 r)H^\prime(r)\\&-(2 M - r) r (35 M + 4 r)H^{\prime\prime}(r)\\& + 4 (2 M - r)^2 r^2 H^{\prime\prime\prime}(r)\big]\Big]Y_{20}(\theta, \varphi)\\&+ 
	\big[r (-60 M^2 + 50 M r - 8 r^2)H(r)\\&-4 (2M-r) r^2 (-3 M + 2 r)H'(r)\big]\partial^2_\theta Y_{20}(\theta, \varphi)\bigg\}\,.
	\end{aligned}
    $}
    \label{crVeff}
\end{equation}

By substituting the profiles $H(r)$ and $K(r)$ that have been determined from Eq.~\eqref{Hout} and Eq.\eqref{KDEex}, calculating the projection of $V_{\rm eff}$ on $Y_{20}(\theta,\varphi)$ by integration, and expanding it at different orders of $r$, we can extract the coefficients of order $r^{-3}$ and $r^2$ that contribute to the NC corrections of the tidal deformability. In this way, the analytical expression of the corrections due to the spacetime non-commutativity of the tidal Love number $k_2$ can be derived to be
\begin{equation}
    k_2 = k_2^{(0)} + k_2^{(2)} \Theta^2\,,
\end{equation}
\begin{equation}
\scalebox{0.85}{$
\begin{aligned}
    k_2^{(2)} =& \frac{M^2}{6 c_1 (2M - R) R^6} \Big[2 c_2 M (M^2 - 6MR + 3 R^2)\\
    &+ M R (2M - R) (K(R)-H(R)) + 6 c_1 R (4M^2 - R^2)\Big]\\
    &+ \frac{c_2 M^2 (M - R)}{2 c_1 R^5} \ln(1 - \frac{2 M}{R})\,.
\end{aligned}
$}
\label{k2(2)}
\end{equation}

Despite the $\ln(1-\frac{2 M}{R})$ term in $k_2^{(2)}$ is divergent near the black-hole-compactness limit $1-\frac{2 M}{R} \rightarrow 0$, we cannot assert that an amplification of $k_2^{(2)}$ in this limit would take place if we do not consider at the same time the behavior of ${c_2}/{c_1}$. By solving $c_1$ and $c_2$ through Eqs.~\eqref{cSolve} while employing Eq.~\eqref{hdef} and Eq.~\eqref{kdef}, the expression of $k_2^{(2)}$ becomes
\begin{equation}
    \scalebox{0.9}{$
    \begin{aligned}
        k_2^{(2)}=&\frac{1}{M^2}C^5\{8C^3(1-2C)k(R)\\
        &+2C(8C^4+20C^3-66C^2+45C-9)h(R)\\
        &+4C(4C^4-8C^3+41C^2-36C+9)\\
        &+9(1-2C)^2[(1-2C)h(R)-2(1-C)]\ln(1-2C)\}\\
        &\{2C(4C^4+6C^3-22C^2+15C-3)h(R)\\
        &+4C(2C^4-2C^3+13C^2-12C+3)\\
        &-3(1-2C)^2[h(R)(1-2C)-2(1-C)]\ln(1-2C)\}^{-1}\,.
    \end{aligned}
    $}
    \label{k2cr_C}
\end{equation}

Since each $\ln(1-2C)$ is multiplied with $(1-2C)^2$, this expression indicates that, as long as the spherical object has non-singular metric on its surface --- so that $h(R)$ and $k(R)$ have finite values --- for a fixed mass $M$ of a celestial compact object, in the black-hole-compactness limit $C\rightarrow\frac{1}{2}$, the leading order NC correction to the tidal Love number $k_2$ will not diverge, but rather it will approach 

\begin{equation}
    \lim_{C\rightarrow\frac{1}{2}, M\,\text{fixed}}k_2^{(2)}=\frac{1}{8M^2}\,.\label{k2cr_BH_L}
\end{equation}

Correspondingly, considering a inspiraling event of a binary system that consists of two identical compact objects with masses $m_1$ and $m_2$, both in the black hole compactness limit, the NC correction \red{to} the GW phase is found to be
\begin{equation}
\lim_{\substack{C_1,C_2\rightarrow\frac{1}{2}, \\m_1,m_2\, \text{fixed}}}\delta\Psi_{TD}^{(NC)}(f)=-\frac{3(\pi f)^\frac{5}{3} (\frac{m_1}{m_2}+\frac{m_2}{m_1}+11)\Theta^2}{2 (m_1+m_2)^\frac{1}{3}}\,.\label{Psicr_BH_L}
\end{equation}
These results indicate that, for a spherical body with non-singular surface, the NC correction to the tidal Love number does not entail, in the black-hole-compactness limit, any localization enhancement to sub-Planckian scale~\cite{Maselli_Horizon_Scale}. Conversely, the precision in the localization of the would-be-horizon 
converges to a finite value, negatively correlated to $M$. Since $k_2^{(0)}$ vanishes in this limit, the existence of a finite limit for $k_2^{(2)}$ further implies that the $k_2^{(2)}\Theta^2$ contribution may dominate the tidal Love number when the compactness approaches $1/2$. In fact, compact objects with smaller masses tend to amplify the NC corrections to the tidal Love number in the black-hole-compactness limit.

\section{Non-Commutative Corrections to Neutron Star Tidal Deformability} \label{4}
\noindent
While a black hole always has vanishing tidal deformability \cite{Binnington_NR_BH_k2_0, Chia_2021_RT_BH_k2_0}, neutron star is the most compact known object that has observable tidal deformability, and it will be the first object of study in this paper.
\label{sec:NC_NS_num}

The structure of a neutron star can be determined thanks to the Tolman-Oppenheimer-Volkoff (TOV) equations \cite{Oppenheimer}, i.e. by solving the system of differential equations 
\begin{equation}
    \begin{aligned}
        &\frac{dm(r)}{dr}=4\pi\rho(r)r^2\,,\\
        &\frac{d\Gamma(r)}{dr}=\frac{m(r)+4\pi r^3 p(r)}{r[r-2m(r)]}\,,\\
        &\frac{p(r)}{dr}=-[p(r)+\rho(r)]\frac{d\Gamma(r)}{dr}\,.
    \end{aligned} \label{TOV}
\end{equation}
where $\rho(r)$ is the mass density, $p(r)$ denotes the pressure, $\rho(0)$ is the mass density value at the origin of the radial coordinates, and the expression $\rho=\rho(p)$ denotes a generic equation of state.

In the internal region of a neutron star, metric perturbations induce a perturbation of the Einstein tensor $G^\mu_\nu=R^\mu_\nu-\frac{1}{2}\delta^\mu_\nu R$ of the form 
\begin{align}
    &\delta G^0_0=-8\pi~\delta\rho~Y_{20}(\theta, \varphi)\,,\\
    &\delta G^1_1=\delta G^2_2=\delta G^3_3=8\pi~\delta p~Y_{20}(\theta, \varphi)\,,\\
    &\delta G^\mu_\nu=0\,,\quad ~\text{for} \quad~\mu\neq\nu \,.\label{GnDia}
\end{align}
If we define $f(p)={d\rho}/{dp}$ according to a given EoS model, and hence consider $\delta\rho=f(p)\delta p$, we may find
\begin{equation}
    \delta G^0_0-\delta G^1_1=-\frac{f(p)+1}{2}(\delta G^2_2+\delta G^3_3)\,. \label{GC}
\end{equation}

According to the functional relation between $G^\mu_\nu$ and $g_{\mu\nu}$, $\delta G^\mu_\nu$ can be expressed in terms of $\Gamma(r), \Lambda(r), H(r)$ and $K(r)$, which encode the NC corrections to $g_{\mu\nu}$ in~Eq.~\eqref{DM}. Thus, Eqs.~\ref{GnDia}- \ref{GC} become 
\begin{align}
    &H''+(\frac{2}{r}+\Gamma'-\Lambda')H'+\Big[-\frac{6e^{2\Lambda}}{r^2}-2(\Gamma')^2+2\Gamma''\nonumber\\&+\frac{3}{r}\Lambda'+\frac{7}{r}\Gamma'-2\Gamma'\Lambda'+\frac{f(p(r))}{r}(\Gamma'+\Lambda')\Big]H=0\,, \label{HDE}
\end{align}
\begin{align}
    K^\prime=H^\prime+2\Gamma^\prime H\,, \label{KDE}
\end{align}
which, according to the substitutions of Eq.~\eqref{hdef} and Eq.~\eqref{kdef}, can be reshuffled into
\begin{equation}
\begin{aligned}
    &r h'+h(h-1)+r (\frac{2}{r}+\Gamma'-\Lambda')h\\
    &+r^2 \Big[-\frac{6e^{2\Lambda}}{r^2}-2(\Gamma')^2+2\Gamma''+\frac{3}{r}\Lambda'\\&+\frac{7}{r}\Gamma'-2\Gamma'\Lambda'+\frac{f(p(r))}{r}(\Gamma'+\Lambda')\Big]=0\,, \label{hDE}
\end{aligned}
\end{equation}
\begin{align}
    rk^\prime=-hk+h+2r\Gamma^\prime \,. \label{kDE}
\end{align}

Boundary conditions for $h(r)$ and $k(r)$ are provided by the $r\rightarrow 0$ limit of Eq.~\eqref{hDE} and Eq.~\eqref{kDE}, i.e.
\begin{align}
    h(0)=&2,\label{hini}\\
    k(0)=&1.\label{kini}
\end{align}

 Thus, we can calculate $k_2^{(2)}$ for different values of the mass of the star, according to Eq.~\eqref{k2(2)} and specific choices of EoS --- namely SLy4 and FPS \cite{Pandharipande1989, Douchin_2001,Haensel_2004} --- we can derive the results shown in FIG.~\ref{fig:SLy4_k2cr} and FIG.~\ref{fig:FPS_k2cr}. 

\begin{figure}[h]
    \centering
    \includegraphics[width=\linewidth]{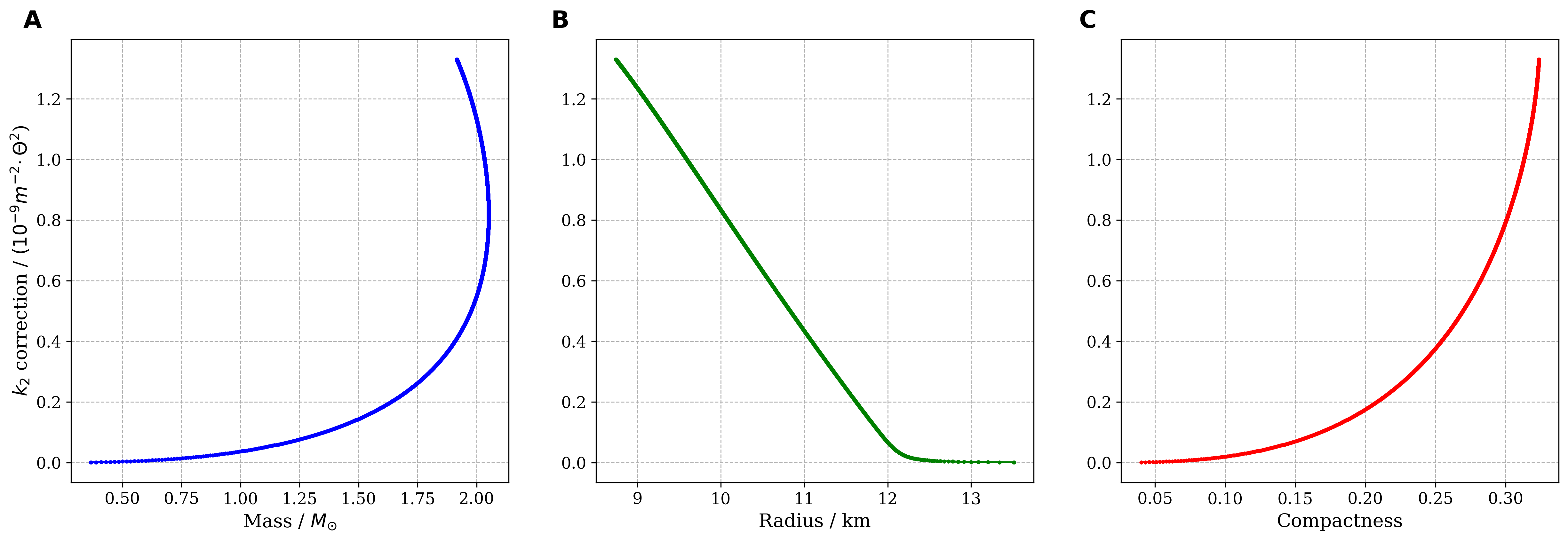}
    \caption{Adopting the EoS SLy4, we plot the leading order spacetime NC correction to the tidal Love number $k_2$ of neutron stars v.s. (A) the neutron star mass $M$, (B) the neutron star radius, (C) the neutron star compactness $C$.}
    \label{fig:SLy4_k2cr}
\end{figure}

\begin{figure}[h]
    \centering
    \includegraphics[width=\linewidth]{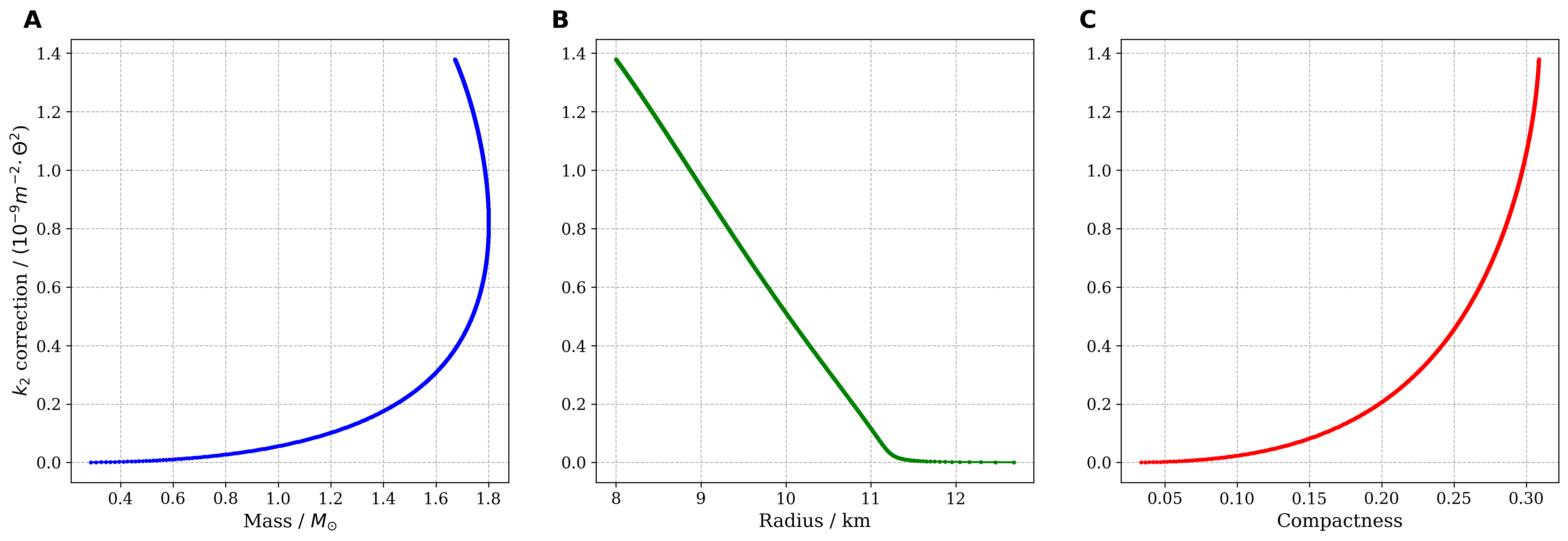}
    \caption{Adopting the EoS FPS, we plot the leading order spacetime NC correction to the tidal Love number $k_2$ of neutrons stars v.s. (A) the neutron star mass $M$, (B) the neutron star radius, (C) the neutron star compactness $C$.}
    \label{fig:FPS_k2cr}
\end{figure}

From these results, we can infer that even for compact objects like neutron stars, the NC correction is difficult to be experimentally distinguished from the contribution due to the unperturbed tidal deformability --- the coefficient in front of $\Theta^2$ in the leading order NC correction is as small as $10^{-9} m^{-2}$. Meanwhile, we can infer that a more compact neutron star has a more significant NC effect on tidal deformability. We can further investigate the imprint of this effect on GW signals, according to Eq.~\eqref{PsiTD}.\\

For neutron stars (which belong to a binary system) with different masses but the same EoS, the NC correction to $\Psi_{TD}$ can be calculated at the leading order, which is the second order in $\Theta$, according to the EoS SLy4 and FPS \cite{Pandharipande1989,Douchin_2001, Haensel_2004}, and at a GW frequency $f=1000\rm{Hz}$. Results are shown in FIG.~\ref{fig:SLy4 dpsi} and FIG.~\ref{fig:FPS dpsi}, from which we notice that in the typical neutron star mass range --- approximately from $1.1\,M_{\odot}$ to $2.2\,M_{\odot}$~\cite{Suwa_2018}\cite{rocha2023massdistributionmaximummass} --- for either SLy4 or FPS, the NC correction to $\Psi_{TD}$ is order $10^{-9} m^{-2} * \Theta^2$, and turns out to be larger for smaller values of the mass of the stars. \\

\begin{figure}[h]
    \centering
    \includegraphics[width=0.6\linewidth]{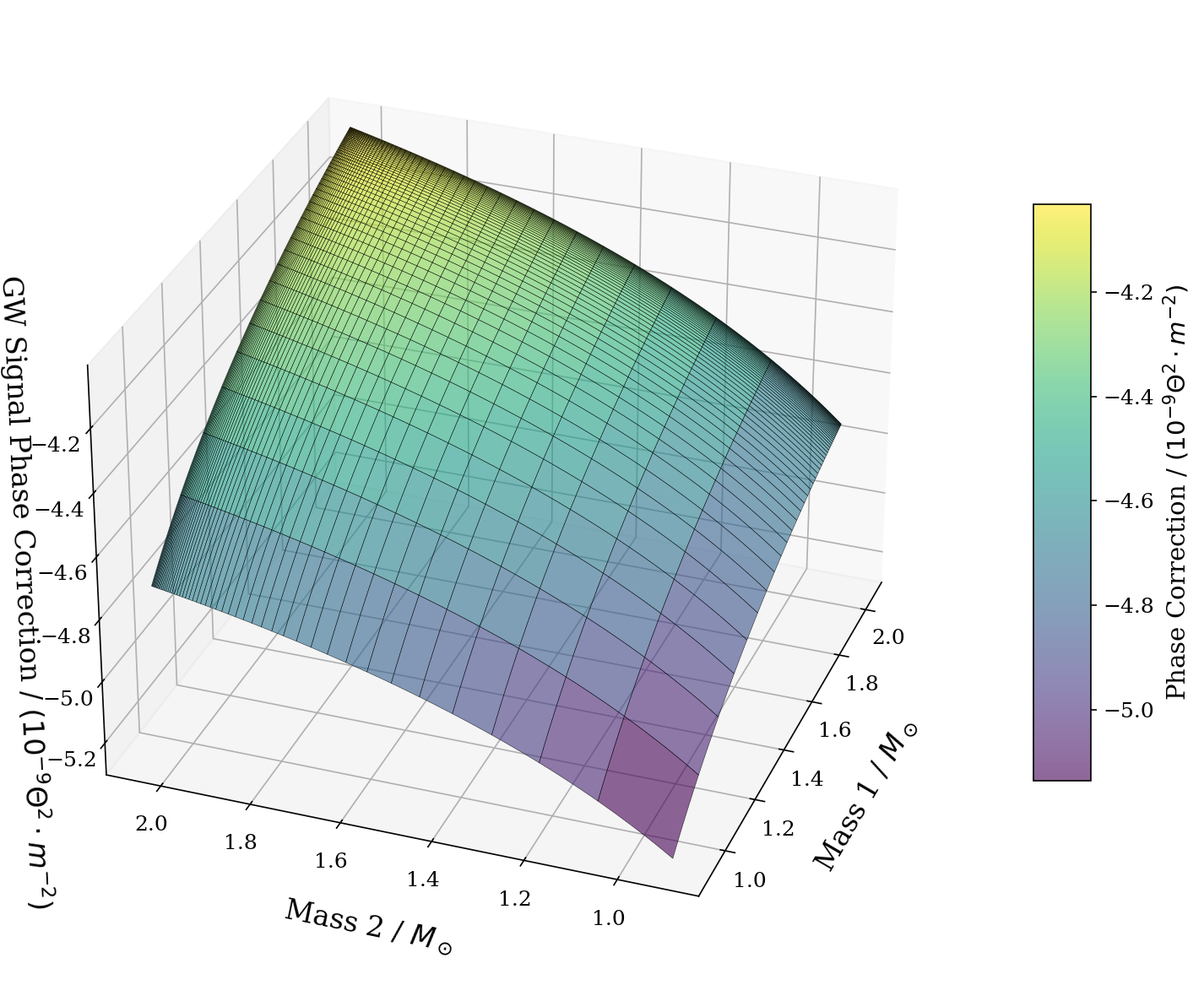}
    \caption{We display the 3D plot of the NC correction to the GW phase $\Psi_{TD}$ at $f=1000\rm{Hz}$ and for different values of the mass of the neutron star binary systems, according to the EoS SLy4.}
    \label{fig:SLy4 dpsi}
\end{figure}

\begin{figure}[h]
    \centering
    \includegraphics[width=0.6\linewidth]{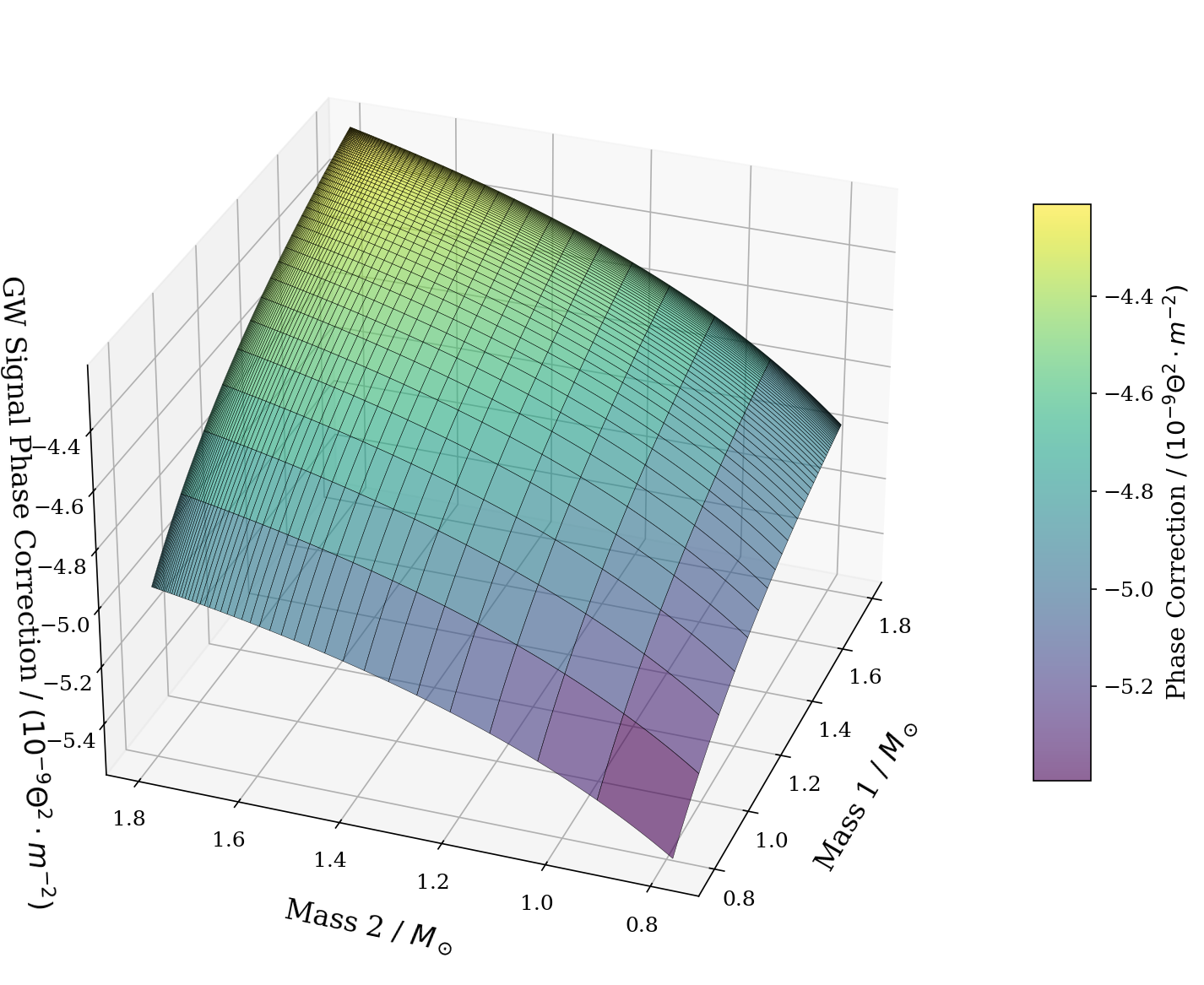}
    \caption{We display the 3D plot of the NC correction to the GW phase $\Psi_{TD}$ at $f=1000\rm{Hz}$ and for different values of the mass of the neutron star binary systems, according to the EoS FPS.}
    \label{fig:FPS dpsi}
\end{figure}
As an example, we may consider a binary system composed of two neutron stars, each one of them with a mass of $1.1\,M_{\odot}$ and satisfying the EoS of SLy4. The absolute value of correction to the GW phase due to tidal deformability can be less than $2.34 \times 10^{-62}$ when $|\Theta|<10^{-11} \, {\rm GeV}^{-1}$ \cite{Kanazawa}, which is far below the precision of current measurement of the GW phase.\\  

However, the increasing trend of the NC effect for less massive neutron stars provides inspiration for better candidates. Any kind of either exotic or not yet observed binary system that would be constituted by stars not as heavy as a neutron star but that would still preserve the potential to emit an observable GW signal in inspiring events, would represent a better candidate to probe spacetime non-commutativity. For these considerations, in the next section we analyse a class of hypothetical exotic compact object, boson stars.

\section{Non-Commutative Corrections to Boson Star Tidal Deformability} \label{5}
\noindent

\subsection{Structure Equations for Boson Star}
\label{sec:CE_BS}
\noindent
Boson stars are exotic compact objects composed by self-repulsively interacting and un-bounded bosons subjected to extremely strong gravity. In order to discuss the NC corrections to boson stars' tidal deformability, we consider as an example a scalar boson field, the action of which reads

\begin{equation}
S = \int d^4x \sqrt{-g} \left[ \frac{R}{16\pi} - g^{ab} \partial_a \Phi^* \partial_b \Phi - V(|\Phi|^2) \right]\,.
\end{equation}

The axion, being a stable particle with self-repulsive interaction, is a suitable candidate to explain the formation of boson stars \cite{Edward_axion_BS}. Its self-interaction potential  \cite{Luca_QCD_axion}\cite{Guerra_2019_axion_BS}\cite{di_Cortona_2016_BS_potential} can be described as
\begin{equation} 
    \scalebox{1.0}{$
    V = \frac{2 \mu_a^2 f_a (m_u + m_d)^2}{m_u m_d} (1 - \sqrt{1 - \frac{4 m_u m_d}{(m_u + m_d)^2}\sin^2 \left( \frac{|\Phi|}{2 f_a} \right)})\,,
    $} \label{V}
\end{equation}
where $m_u$ and $m_d$ are the masses of the up quark and the down quark, respectively, while $f_a$ is a scaling constant \cite{di_Cortona_2016_BS_potential} constrained to be in the range from $10^8$ GeV to $10^{17}$ GeV \cite{Raffelt2008_f_a, Arvanitaki_f_a_1, Arvanitaki_f_a_2, Arvanitaki_f_a_3}. In Eq.~\eqref{V} $\mu_a$ denotes the mass of the axion, determined by \cite{di_Cortona_2016_BS_potential}
\begin{equation}
    \mu_a \approx 5.7*10^{-6} \,\text{eV} \left( \frac{10^{12} \text{ GeV}}{f_a} \right)\,,
\end{equation}

We assume to expand the potential $V$ up to the fourth order $|\Phi|^4$, i.e.
\begin{equation}
    \scalebox{0.95}{$
    V(|\Phi|^2)= \mu_a^2|\Phi|^2 + (-\frac{1}{12} + \frac{1}{4} B)\frac{\mu_a^2}{f_a^2}|\Phi|^4+O(|\Phi|^6)\,,
    $}
\end{equation}
where $B = {m_u m_d}/{(m_u + m_d)^2}$, and $B$ can be determined from the ratio $m_u / m_d \approx 0.48$ \cite{ZDRAHAL201168_quark_mass}.\\

To solve for the boson star structure semi-classically, we adopt the ansatz \cite{Liebling_2023}
\begin{equation}
\Phi(\vec{r}, t) = \phi_0(r) \e^{-i \omega t}\,.
\end{equation}

We can derive the equations of motion by varying the action with respect to the metric variables $g_{\mu\nu}$ (Einstein field equations), and the scalar field $\Phi$ along with its complex conjugate $\Phi^*$ (Klein-Gordon equation). The Einstein field equations reduce to
\begin{equation}
    \scalebox{0.9}{$
    \Lambda'(r) = \frac{1 - \e^{2 \Lambda}}{2 r} + 4\pi r \left( \omega^2 \e^{2 \Lambda - 2 \Gamma} \phi_0^2 + \phi_0'^2 + \e^{2 \Lambda} V(|\Phi|^2) \right)\,, 
    $}
    \label{eqlambda}
\end{equation}
\begin{equation}
    \scalebox{0.9}{$
    \Gamma'(r) = \frac{\e^{2 \Lambda} - 1}{2 r} + 4\pi r \left( \omega^2 \e^{2 \Lambda - 2 \Gamma} \phi_0^2 + \phi_0'^2 - \e^{2 \Lambda} V(|\Phi|^2) \right)\,.
    $}
    \label{eqgamma}
\end{equation}

while the Klein-Gordon Equation reads:
\begin{align}
    \scalebox{0.9}{$
    \phi_0''(r) = e^{2 \Lambda} \left( \frac{d V(|\Phi|^2)}{d(|\Phi|^2)} - \omega^2 e^{-2 \Gamma} \right) \phi_0 
    + \left( \Lambda' - \Gamma' - \frac{2}{r} \right) \phi_0'\,.
    $}
    \label{eqKG}
\end{align}
Eqs.~\eqref{eqlambda}-\eqref{eqKG} determine the structure of the boson star in a similar way as the TOV equations, with the mass density and the pressure being replaced by

\begin{align}
    \rho(r) &= \omega^2 \e^{- 2 \Gamma} \phi_0^2 + \e^{2 \Lambda} \phi_0'^2 + V(|\Phi|^2)\,, \\
    p(r) &= \omega^2 \e^{- 2 \Gamma} \phi_0^2 + \e^{2 \Lambda} \phi_0'^2 - V(|\Phi|^2)\,.
\end{align}

Despite the similarity between the Boson star's structure equations and the TOV equations for a neutron star, solving the former ones can be much more tricky.
Eqs.~\eqref{eqlambda}-\eqref{eqKG} appear as an eigenvalue problem in $\omega$.  The values of $\Lambda'(r)$ and $\phi_0'(r)$ at the center of the star must vanish, due to spherical symmetry. This leads to a vanishing value at the center of the star also for $\Lambda(r)$, due to Eq.~\eqref{eqlambda}. Furthermore, the value of $\Gamma$ at the center of the star can be set arbitrarily and tuned by varying $\omega$. In fact, $\Gamma$ must be added by a constant in order to vanish at infinity while keeping $\omega^2 \e^{-2 \Gamma(0)}$ invariant in the last step, because the metric must be asymptotically flat. Thus the only two variables left to determine the solution of Eqs.~\eqref{eqlambda}-\eqref{eqKG} are the central value of $\phi_0(r)$ (denoted by $\phi_0^{(c)}$) and $\omega$.\\

For each $\phi_0^{(c)}$, which yield different boson star structures, $\omega$ is fixed by requiring $\phi_0(r) \to 0$ as $r \to \infty$, due to requirements of physicality and asymptotic flatness. For every $\omega$ that satisfies this constraint, we will select the smallest value that provides the most stable structure of the boson star \cite{Stable_BS}. Due to the properties of these differential equations, even small deviations from those specific values of $\omega$ (especially negative deviations) that satisfy the constraint may lead to divergences in the solution at spatial infinity. To determine the values of $\omega$, we developed an algorithm that tunes $\omega$ in order to minimize $|\phi_0'(r_0)|$ --- $r_0$ denotes where $\phi_0(r_0) \approx 0$, within $10^{-6}$ accuracy. 

In the following numerical evaluations, we have selected as an example the value $f_a = 10^8$ GeV. 

\subsection{Probing Non-Commutative Effects via Boson Star}
\label{sec:NC_BS_num}
\noindent
Upon solving $\Lambda(r)$, $\Gamma(r)$ and $\phi_0(r)$, we can similarly consider the tidal perturbation on metric as shown in Eq.~\eqref{DM}. The corresponding perturbation in the matter field will be reflected as a perturbation term on the boson field $\Phi(\vec{r}, t)$, namely

\begin{equation}
    \delta\Phi(\vec{r}, t) = \phi_1(r) \e^{-i \omega t} Y_{20}(\theta, \varphi)\,.
\end{equation}

Considering the first order tidal perturbation on the Einstein field equations and the Klein-Gorden equation, we will get:

\begin{equation}
\scalebox{0.9}{$
\begin{aligned}
    &H'' + \left(\frac{2}{r} + \Gamma' - \Lambda' \right) H' \\
    &- \left[ 2 \Gamma'^2 - \frac{4 \Gamma'}{r} - 48\pi \e^{2 \Lambda - 2 \Gamma} \omega^2 \phi_0^2 + 16\pi \phi_0'^2 
    + \frac{8 \e^{2 \Lambda} - 2}{r^2} \right] H \\
    &+ 32\pi \left[ \phi_0'' - \left( \Lambda' + \Gamma' - \frac{2}{r} \right) \phi_0' - \omega^2 \phi_0 \e^{2 \Lambda - 2 \Gamma} \right] \phi_1 = 0\,,
\end{aligned}
$}
\label{H_BS}
\end{equation}
\begin{equation}
\scalebox{0.9}{$
\begin{aligned}
    &\phi_1'' + \left(\frac{2}{r} + \Gamma' - \Lambda' \right) \phi_1' \\
    &- \biggl[ \e^{2 \Lambda} \left( \frac{d V(|\Phi|^2)}{d(|\Phi|^2)} + 2 \frac{d^2 V(|\Phi|^2)}{[d(|\Phi|^2)]^2} \phi_0^2 - \omega^2 e^{-2 \Gamma} \right)\\
    &+ 32\pi \phi_0'^2 + \frac{6 \e^{2 \Lambda}}{r^2} \biggr] \phi_1 \\
    &+ \biggl[ \phi_0'' - \left( \Lambda' + \Gamma' - \frac{2}{r} \right) \phi_0' - \omega^2 \phi_0 \e^{2 \Lambda - 2 \Gamma} \biggr] H = 0\,,
\end{aligned}
$}
\label{Phi1_BS}
\end{equation}
\begin{equation}
    K' = H' + 2 H \Gamma' + 32 \pi \phi_1\phi_0''\,.
    \label{K_BS}
\end{equation}

We can now solve for $H(r)$, and calculate the Boson stars' tidal deformability and the corrections due to spacetime non-commutativity. At this purpose, we employ relations Eqs.~\eqref{hdef}-\eqref{kdef}, and further define
\begin{align}
\Phi_1(r)&:=\frac{r \phi_1'(r)}{\phi(r)}\,,\\
\psi_1(r)&:=\frac{\phi_1(r)}{H(r)}
=C\exp\Big[\int\frac{\Phi_1(r)-h(r)}{r}\d r\Big]\,,\label{psi_1_def}
\end{align}
where C is a constant. Eqs. \eqref{H_BS}-\eqref{K_BS} thus become
\begin{equation}
\scalebox{0.9}{$
\begin{aligned}
    &r h'+h(h-1)+r (\frac{2}{r}+\Gamma'-\Lambda')h\\
    &-r^2 \left[ 2 \Gamma'^2 - \frac{4 \Gamma'}{r} - 48\pi \e^{2 \Lambda - 2 \Gamma} \omega^2 \phi_0^2 + 16\pi \phi_0'^2 
    + \frac{8 \e^{2 \Lambda} - 2}{r^2} \right]\\
    &+ 32\pi r^2 \left[ \phi_0'' - \left( \Lambda' + \Gamma' - \frac{2}{r} \right) \phi_0' - \omega^2 \phi_0 \e^{2 \Lambda - 2 \Gamma} \right] \psi_1 =0\,, 
\end{aligned}
$}
\end{equation}
\begin{equation}
\scalebox{0.9}{$
\begin{aligned}
    &r\Phi_1' + \Phi_1(\Phi_1-1) + r \left(\frac{2}{r} + \Gamma' - \Lambda' \right) \Phi_1 \\
    &- \biggl[ \e^{2 \Lambda} \left( \frac{d V(|\Phi|^2)}{d(|\Phi|^2)} + 2 \frac{d^2 V(|\Phi|^2)}{[d(|\Phi|^2)]^2} \phi_0^2 - \omega^2 e^{-2 \Gamma} \right)\\
    &+ 32\pi \phi_0'^2 + \frac{6 \e^{2 \Lambda}}{r^2} \biggr] \\
    &+ \biggl[ \phi_0'' - \left( \Lambda' + \Gamma' - \frac{2}{r} \right) \phi_0' - \omega^2 \phi_0 \e^{2 \Lambda - 2 \Gamma} \biggr] \psi_1^{-1} = 0\,,
\end{aligned}
$}
\end{equation}
\begin{equation}
\begin{aligned}
    rk^\prime=-hk+h+2r\Gamma^\prime +32\pi r \phi_0''\psi_1\,. 
\end{aligned}
\end{equation}
From these differential equations we derive the boundary conditions $h(0)=2$ and $\Phi_1(0)=2$, while $\psi_1(r)$ will be a non-vanishing function on $[0, +\infty]$, according to Eq.~\eqref{psi_1_def}. 
From these boundary conditions follow quadratic boundary conditions for $H$ and $\phi_1$ near the star center, i.e.

\begin{align} 
    &H(r) = H^{(c)} r^2 + O(r^3)\,,\\
    &\phi_1 = \phi_1^{(c)} r^2 + O(r^3)\,.
\end{align}

Solving Eq.~\eqref{H_BS} and Eq.~\eqref{Phi1_BS} is a similar task to the one we discussed in Section \ref{sec:CE_BS}. The value of $H^{(c)}$ can be chosen arbitrarily, inducing a scaling of both $H(r)$ and $\phi_1$, without affecting the final value of tidal deformability, while $\phi_1^{(c)}$ shall be adjusted in order to ensure that $\phi_1$ converges to $0$ at spatial infinity. 


Adopting a similar methodology to the one discussed in the previous sections, we may calculate the tidal deformability and its leading order NC correction --- results are shown in Figure~\ref{fig:k2_BS} and Figure~\ref{fig:k2cr_BS}.
\begin{figure}[h]
    \centering
    \includegraphics[width=\linewidth]{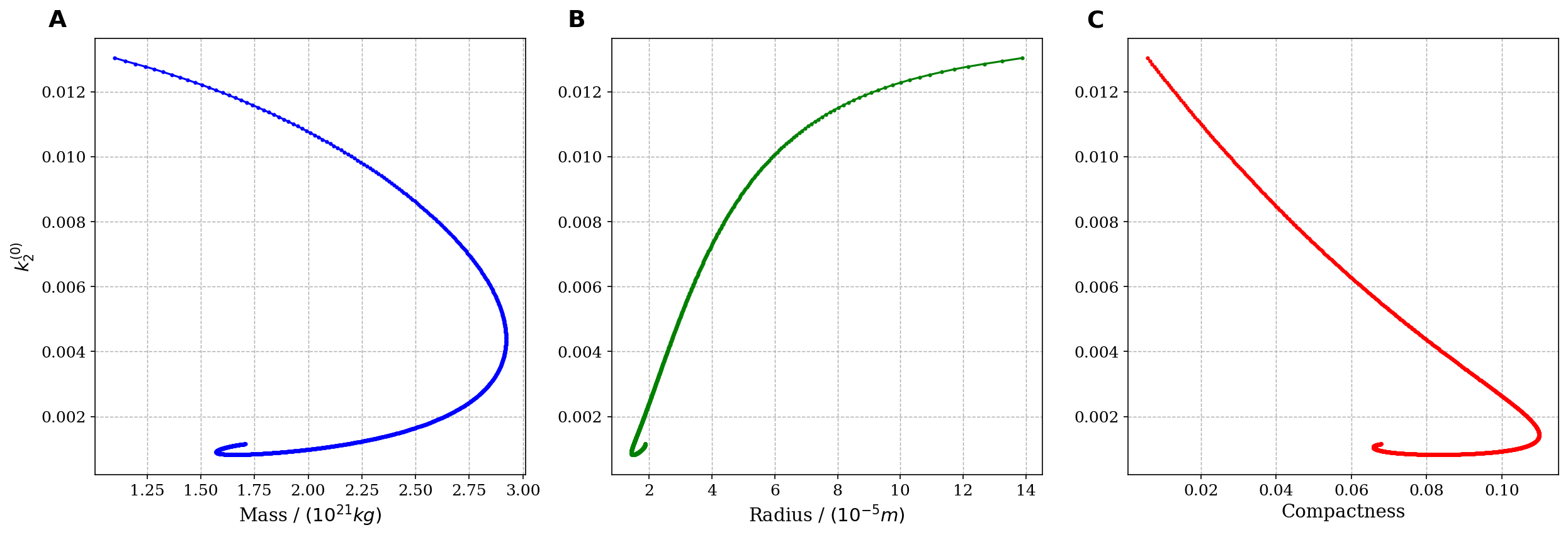}
    \caption{Fixing $f_a = 10^8$ GeV, we display the tidal Love number $k_2$ of boson stars v.s. (A) the boson star mass $M$, (B) the boson star radius, (C) the boson star compactness $C$.}
    \label{fig:k2_BS}
\end{figure}
\begin{figure}[h]
    \centering
    \includegraphics[width=\linewidth]{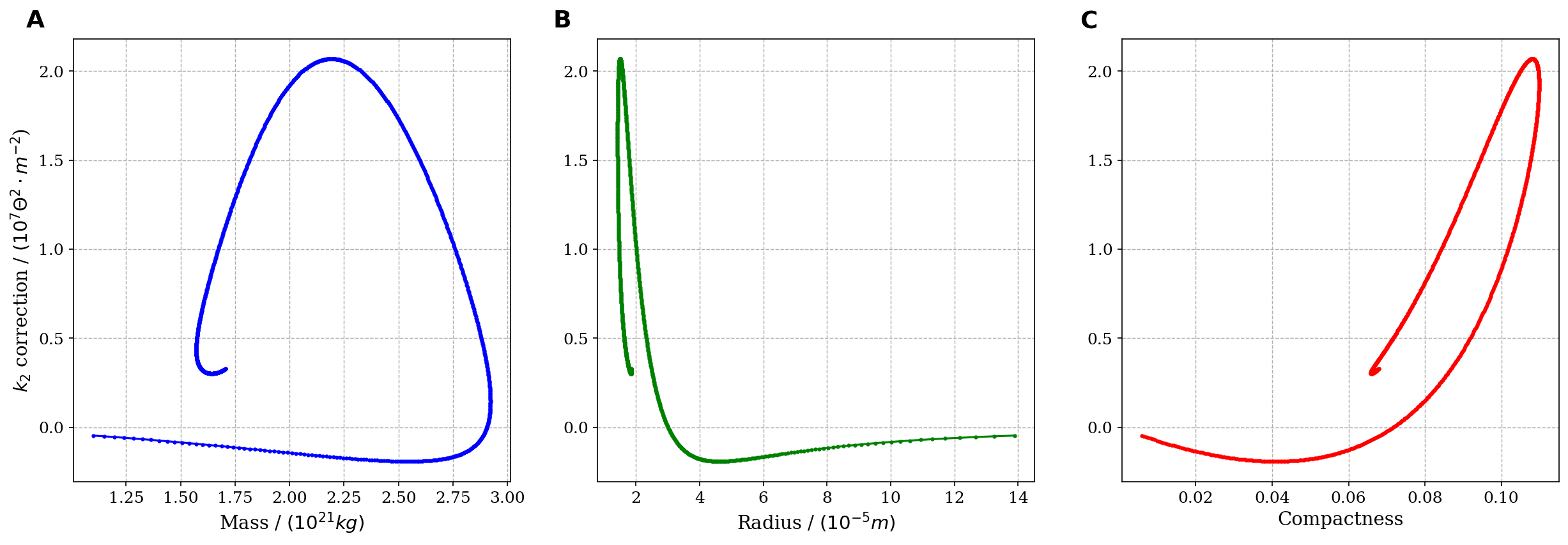}
    \caption{Fixing $f_a = 10^8$ GeV, we display the leading order NC correction to the tidal Love number $k_2$ of boson stars v.s. (A) the boson star mass $M$, (B) the boson star radius, (C) the boson star compactness $C$.}
    \label{fig:k2cr_BS}
\end{figure}
We may also calculate the corresponding NC correction (at  leading order) to the phase of the GW emitted by the binary systems, considering different values of the mass and of the compactness of the inspiraling boson stars --- results are shown in Figure~\ref{fig:GW_B_BS}, having fixed $f = 1000\text{Hz}$ (default choice in this section).\\

\begin{figure}[h]
    \centering
    \includegraphics[width=\linewidth]{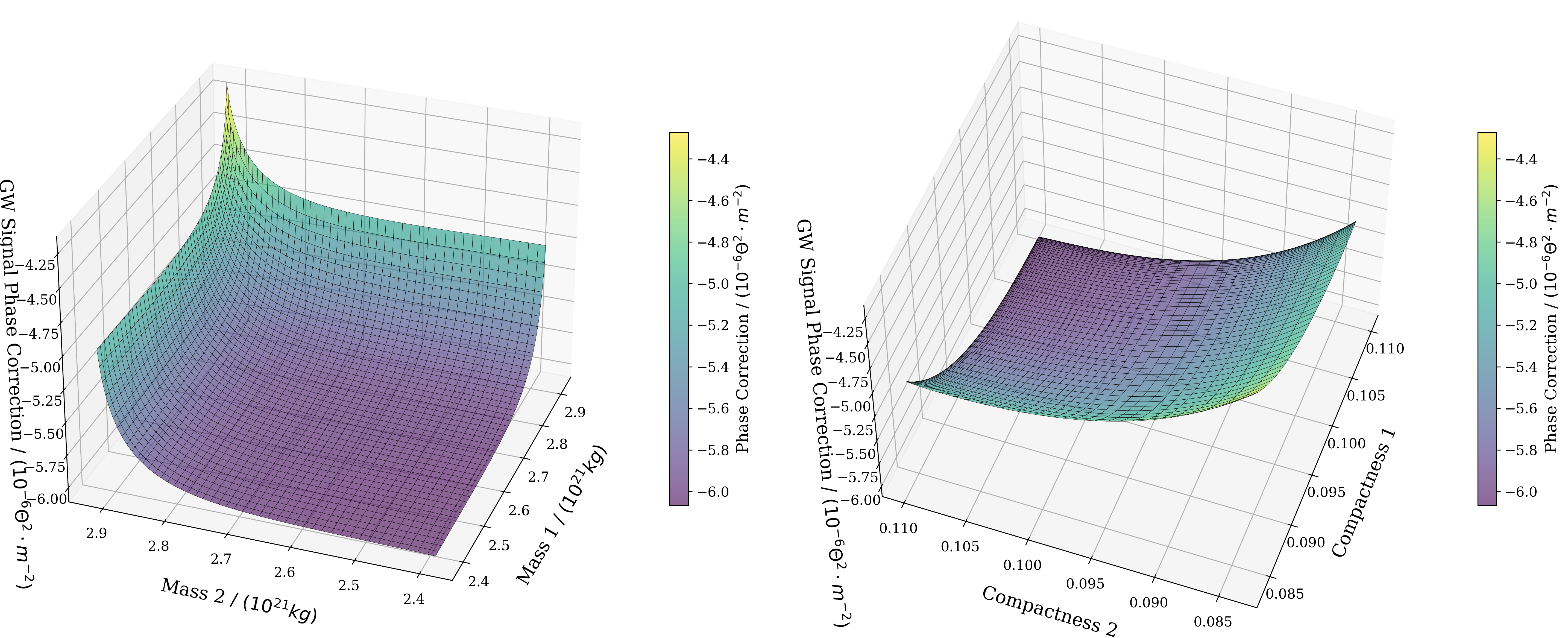}
    \caption{We display the 3D plot of the NC correction to the GW phase $\Psi_{TD}$ at $f=1000\rm{Hz}$, for different values of the mass (left panel, SI units are used) and compactness (right panel, Planck units are used) of the two boson star in the binary systems, having fixed $f_a = 10^8$ GeV for both the inspiraling boson stars.}
    \label{fig:GW_B_BS}
\end{figure}
At leading order, for the boson star binary systems under scrutiny, the magnitude of the NC corrections to the observable quantities can be summarized to be 
\begin{align}
    \delta k_2^{(NC)}\sim&10^{7}\Theta^2 \text{m}^{-2}\,, & \\
    |\delta\Psi_{TD}^{(NC)}|\sim& 10^{-6}\Theta^2 \text{m}^{-2}\,,\label{dpsiTDex1}
\end{align}
where $k_2^{(NC)}=k_2^{(2)}\Theta^2$ represents the leading order magnitude of the NC correction to the tidal Love number, and $\delta\Psi_{TD}^{(NC)}$ denotes the leading order NC correction to $\Psi_{TD}$.\\

We may compare these results with those ones obtained considering the neutron star case, for which we found instead that
\begin{align}
    \delta k_2^{(NC)}\sim&10^{-9}\Theta^2 \text{m}^{-2}\,, & \\
    |\delta\Psi_{TD}^{(NC)}|\sim&10^{-9}\Theta^2 \text{m}^{-2}\,.
\end{align}
\\

NC corrections to the tidal deformability and to the GW phase for boson star binary systems are in magnitude manifestly more significant than those for neutron stars binary systems. Nonetheless, these are still very small if $|\Theta|$ is assumed to acquire values (justified by theoretical arguments previously discussed) that are smaller than $10^{-11}\text{GeV}^{-1}\approx1.973\times10^{-27}\text{m}$ \cite{Kanazawa}.\\ 

Let us consider now compact objects with mass in the range $5.53 \times 10^{-10} \rm{M_\odot} \, \div \, 1.47\times 10^{-9}  \rm{M_\odot}$ --- these are typical values of the masses of the boson stars under scrutiny. If these objects can be compressed up to a compactness close to the one of a black hole, the compactness limit of $\delta k_2^{(NC)}$ can be estimated by Eq.~\eqref{k2cr_BH_L} to be
\begin{equation}
    \begin{aligned}
    \lim_{C\rightarrow\frac{1}{2}, M\,\text{fixed}}\delta k_2^{(NC)}
    \sim &10^{11}\Theta^2 m^{-2}\,.
    \end{aligned}
\end{equation}

For a binary inspiral consisting of two compact objects, both with typical boson star masses, the limit of $\delta\Psi_{TD}^{(NC)}$ can also be estimated, according to~\eqref{Psicr_BH_L}, and found to be
\begin{equation}
    \begin{aligned}
    \lim_{C_1,C_2\rightarrow\frac{1}{2}}|\delta\Psi_{TD}^{(NC)}|\sim10^{-6}\Theta^2 m^{-2}\,.
    \end{aligned}
\end{equation}

NC corrections \red{to} $k_2$ get amplified in this limit, while NC corrections to $\Psi_{TD}^{(NC)}$ remain of the same order of magnitude of typical boson stars. This situation can change a bit if we consider a higher mass ratio of the two stars in a binary system. Assuming the binary system to consist of one neutron star (with EoS SLy4) and one boson star (with boson star model previously specified), as an example, the $\delta\Psi_{TD}^{(NC)}$ for different values of the mass and of the compactness of the two stars is shown in Figure~\ref{fig:GW_NS_BS}.\\

\begin{figure}[h]
    \centering
    \includegraphics[width=\linewidth]{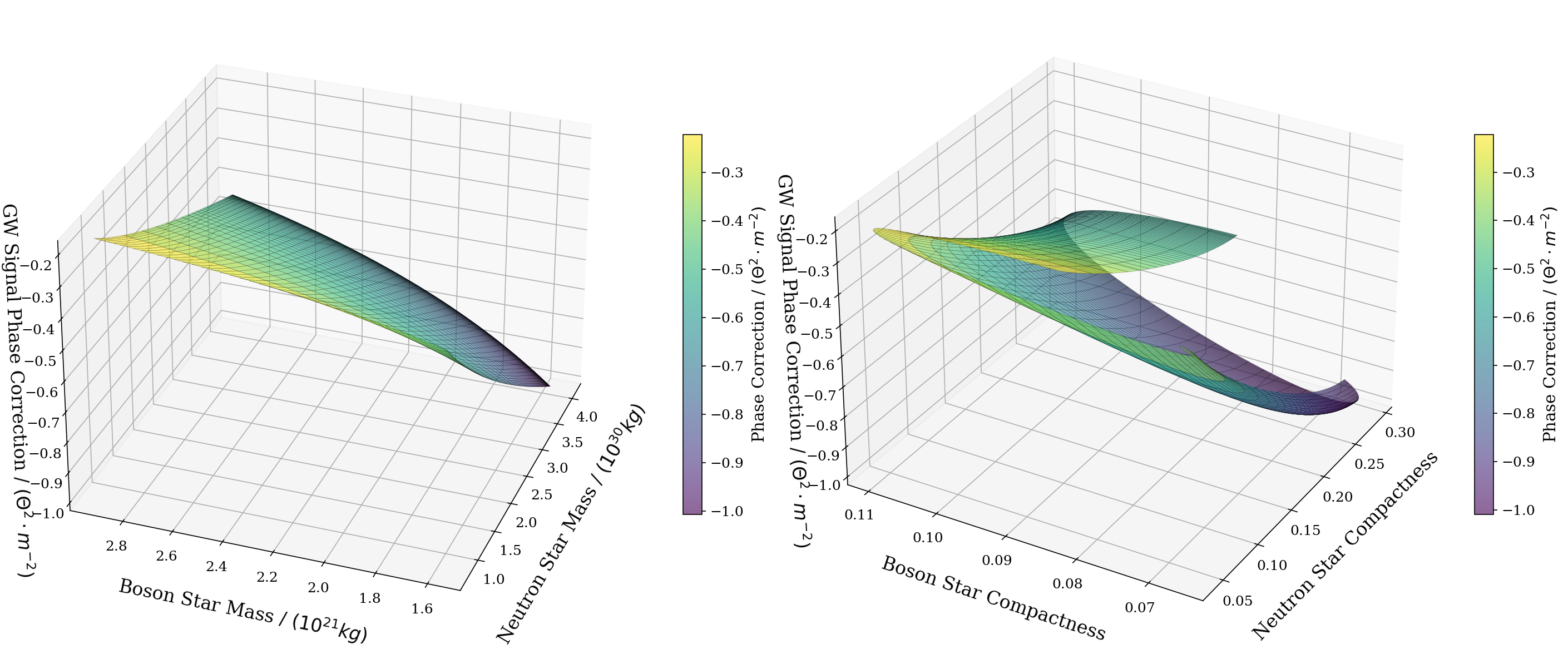}
    \caption{We display the 3D plot of the NC corrections to the GW phase $\Psi_{TD}$ at $f=1000\rm{Hz}$, for different values of the mass (left panel, SI units are used) and of the compactness (right panel, Planck units are used) of a binary system composed by a neutron star and a boson star, with $f_a = 10^8$ GeV for the latter.}
    \label{fig:GW_NS_BS}
\end{figure}

In this binary system model, $|\delta\Psi_{TD}^{(NC)}|$ may acquire values between $0.17\Theta^2m^{-2}$ and $1.45\Theta^2m^{-2}$. Within the black-hole-compactness limit of the two component stars, $|\delta\Psi_{TD}^{(NC)}|$ is in the range between $0.23\Theta^2m^{-2}$ and $1.93\Theta^2m^{-2}$. In fact, in this limit the high mass ratio contributes to $|\delta\Psi_{TD}^{(NC)}|$ manifestly, according to Eq.~\eqref{Psicr_BH_L}. All the values of $|\delta\Psi_{TD}^{(NC)}|$ are larger than the ones reachable in neutron star binaries and boson star binaries. In this mixed case, in fact, the order of magnitude of $|\delta\Psi_{TD}^{(NC)}|$ is mainly determined by the tidal Love number of the neutron star, since its radius, much larger than the one of the companion boson star, renders its tidal deformability dominant in Eq.~\eqref{PsiTD}.\\ 

This suggests that $|\delta\Psi_{TD}^{(NC)}|$ can be further amplified when we consider a boson star, which has a $k_2^{(2)}$ much more sizable than typical neutron stars (participating in a binary inspiraling event together with an even smaller compact object). As a toy model, if we consider a binary inspiral consisting of a boson star described in the previous model and an exotic compact object weighing only $1$kg, compressed to the black-hole-compactness limit, we may find that, while the $\delta k_2^{(NC)}$ of this $1\rm{kg}$ compact object approaches $10^{53}\Theta^2m^{-2}$ --- relatively sizable even for $\Theta$ of the order of magnitude of $10^{-27}\rm{m}$ --- $|\delta\Psi_{TD}^{(NC)}|$ is within the range between $10^{18}\Theta^2 m^{-2}$ and $10^{24}\Theta^2 m^{-2}$. These values cannot yet ensure that NC corrections are observable, since it would be required to have a $\Theta$ as large as $10^{-13}m$ in order to induce a fluctuation in $\Psi_{TD}$ of 0.01 even for the largest estimated value of $|\delta\Psi_{TD}^{(NC)}|$. \\

How physically reliable are the circumstances under which the novel structures of the exotic compact objects we considered induce significant NC corrections to the tidal deformability? We postpone this important question to forthcoming analyses, in which we will further investigate both the amplification mechanism and the implementation of the black-hole-compactness limit, responsible for the vanishing of $k_2^{(0)}$. We notice anyway that the GW phase was recovered not to encode any sizable NC correction, at least relying on the binary configurations we investigated. These configurations are indeed characterized by the convergence of $k_2^{(2)}$ in the black-hole-compactness limit, as discussed in Section \ref{sec:k2cr_NS} and clarified by numerical results in Section \ref{sec:NC_NS_num} and Section \ref{sec:NC_BS_num}.

\section{Conclusions} \label{6}
\noindent
We have first reviewed a consistent theoretical framework in which a de Sitter gauge theory of gravity \cite{Zet} incorporates spacetime non-commutativity effects via the Seiberg-Witten map \cite{Seiberg_1999}. This framework served as the foundation for calculating non-commutative (NC) corrections to the tidal deformability of compact objects and analyzing their imprint on gravitational wave (GW) signals \cite{Hinderer_2010, Hinderer_2008, Flanagan_2008}. From this starting point, we derived an analytical expression for the NC correction to tidal deformability and demonstrated its convergence in the black-hole-compactness limit.

We then applied this analysis to two types of objects: neutron stars, the most compact known astrophysical objects, and boson stars, a hypothetical class of exotic compact objects. For neutron stars, we examined two equations of state \cite{Pandharipande1989, Douchin_2001, Haensel_2004}, in order to compute the NC corrections to tidal deformability and their impact on the GW phase, and presented results for different neutron star masses. For boson stars, we employed a semi-classical approach based on field theory \cite{Cardoso_TL_BS, Mecedo_BS, Guerra_2019_axion_BS}, in order to determine their structures, followed by a similar analysis of NC corrections to the tidal deformability and the GW wave phase. The scale of these corrections was compared to neutron stars, boson stars, and mixed binary systems. Our results indicate that NC corrections to the tidal Love number tend to be amplified by denser objects, and observable GW effects are enhanced by larger mass ratios. In addition, we propose that while the leading-order NC corrections can become dominant for the tidal deformability as the compactness approaches the black-hole limit, neutron stars and boson stars remain inadequate candidates for detecting NC effects via GW observations based on tidal deformability. In fact, while both $k_2$ and $\Psi_{TD}$ are dimensionless, the NC correction to $\Psi_{TD}$ for a binary merger event appears less significant than the NC corrections to the $k_2$ of the compact object components, within the case of extreme compactness. These results underscore the need to further investigate novel exotic compact objects, and motivate the exploration of more direct measurements of the dimensionless tidal deformability.

\vspace{0.3cm}

\bibliographystyle{apsrev4-2}
\bibliography{references}

\end{document}